\documentclass[aps,pre,reprint, amsmath, amssymb,superscriptaddress]{revtex4-1}
\usepackage{morefloats}
\usepackage{bm}
\newcommand{\beq}{\begin{equation}}
\newcommand{\eeq}{\end{equation}}

\usepackage[retainorgcmds]{IEEEtrantools}
\usepackage{graphicx,tikz,placeins}
\usepackage{mathrsfs}
\usepackage{amsmath,amssymb,amsfonts,physics}
\usepackage{color}
\usepackage{float}
\usepackage{times,txfonts}
\usepackage{nicefrac}
\usepackage{ulem}
\usepackage[colorlinks=true,linkcolor=blue,urlcolor=blue,citecolor=blue,pdfusetitle]{hyperref}
\usepackage{physics}
\usepackage{soul}
\usepackage{graphicx,tikz,placeins}
\usepackage{mathrsfs}
\usepackage{amsmath,amssymb,amsfonts,physics}
\usepackage{color}
\usepackage{float}
\usepackage{times,txfonts}
\usepackage{nicefrac}
\usepackage{ulem}
\usepackage[colorlinks=true,linkcolor=blue,urlcolor=blue,citecolor=blue,pdfusetitle]{hyperref}
\usepackage{physics}
\usepackage{soul}
\usepackage[utf8]{inputenc}

\usepackage{cancel}
\newcommand{\GF}[1]{\textcolor{black}{#1}}

\begin{document}

\title{\GF{Optimal Control Strategy for Collisional Brownian Engines}}

\author{Gustavo A. L. For\~ao}
\affiliation{Universidade de São Paulo,
Instituto de Física,
Rua do Matão, 1371, 05508-090
São Paulo, SP, Brazil}

\date{\today}

\begin{abstract}
Collisional Brownian engines have recently gained attention as alternatives to conventional nanoscale engines. However, a comprehensive optimization of their performance, which could serve as a benchmark for future engine designs, is still lacking. In this work, \GF{we build upon this} by deriving and analyzing the optimal control strategy for a collisional Brownian engine. By maximizing the average output work, we show that the optimal strategy consists of linear force segments separated by impulsive delta-like kicks that instantaneously reverse the particle’s velocity. This structure enforces constant velocity within each stroke, enabling fully analytical expressions for optimal output power, efficiency, and entropy production. We demonstrate that the optimal strategy significantly outperforms standard ones (such as constant, linear, or periodic drivings), achieving higher performance while keeping entropy production under control. Remarkably, when evaluated using realistic experimental parameters, the efficiency approaches near unity at the power optimum, with entropy production remaining well controlled. To analyze a more realistic scenario, we examine the impact of smoothing the delta-like forces by introducing a finite duration and find that, although this reduces efficiency and increases entropy production, the optimal strategy still delivers high power output in a robust manner.
\end{abstract}

\maketitle

\section{Introduction}

The study of small-scale engines operating far from equilibrium has gained significant attention in recent years \cite{sekimoto2010stochastic,jarzynski,seifert2012stochastic,VANDENBROECK20156,landientropy}. This interest is particularly driven by their relevance in systems where thermal fluctuations are no longer negligible, such as colloidal particles \cite{martinez2016brownian, albay2021shift,singlestochastig,expbrowargun,Duet,collreview,ciliberto2,Blickle2012}, biomolecular motors \cite{activecoll,ritortfelix,holubec2020active,holubecunder,liepelt2007kinesin,liepelt2009operation,losrios} and quantum dots \cite{subrt_2007,qdesposito,fernandoqd,twocoupled,Harunari_2020}. In these regimes, classical thermodynamic approaches fall short, and a more refined description is offered by stochastic thermodynamics, which captures the role of fluctuations, dissipation, and energy conversion at the microscopic level. One of the main challenges in this context is to extract useful work in finite time while managing entropy production—an issue that has been addressed through a variety of proposed setups, including, for example, collective systems \cite{Vroylandt_2017,collective1,collective2,collective3,mamede} and driving \GF{strategies} tailored to optimize either efficiency or power \cite{curzonahl,seifert2007,seifert2008,imparato1,filho1,mamede2022,harunariasy,angel,Noa1}.

Among them, Brownian engines stand out as both theoretically rich and experimentally accessible platforms, capable of shedding light on the fundamental limits of performance under non-equilibrium conditions. These systems, typically realized with colloidal particles in time-dependent potentials \cite{martinez2016brownian,albay2021shift,singlestochastig,Blickle2012} or optically driven traps \cite{Duet}, provide precise control and measurement of work, heat, and fluctuations, making them ideal candidates for exploring how microscopic dynamics translate into thermodynamic behavior. In such setups, a crucial aspect is the driving \GF{strategy} \GF{(often referred to as the `driving protocol')}—the way in which external forces are applied to the system, shaping its interaction with the environment. One particularly well-studied case is that of engines driven by stiffness modulation, for which optimal driving \GF{strategies} have been determined in both passive and, more recently, active scenarios \cite{seifert2007,seifert2008,sivak,sarah1,sarah2,olsen2025,bonança1,bonança2,naze2022optimal,chatterjee,automaticd,adri,whitelam,aurell}, setting useful performance benchmarks and deepening our understanding of control at small scales. In parallel, focus has shifted to collisional Brownian engines, where a particle confined in a static potential is driven by external forces that directly alter its momentum. Many studies have explored ways to optimize collisional engines' performance by carefully tuning the driving \GF{strategy} \cite{Holubec_2014,filho1,mamede2022,mamede,forao2025,proesmans2017underdamped,PhysRevLett.115.090601,harunariasy,proesmans2016,angel,Noa1}, shedding light on key aspects of protocol design for these systems. Despite this progress, however, a universally optimal \GF{strategy} (hereafter OS)  that serves as an upper limit for the performance of such systems remains elusive.

In this paper, we aim to address this gap by \GF{establishing a theoretical limit for the performance of} collisional Brownian engines, focusing on systems where a single particle interacts sequentially with a thermal reservoir and is subjected to a specific work source at each stage. To \GF{mathematically find} the OS, we use the ideas of Refs. \cite{seifert2007,seifert2008} for the most general case of a collisional Brownian engine described by an underdamped Langevin equation subjected to an external driving force. From this framework, we derive an exact expression for the OS, which includes delta jumps as a key feature, a hallmark also seen in stiffness modulation protocols \cite{seifert2007,seifert2008,sivak,sarah1,sarah2,olsen2025,bonança1,bonança2,naze2022optimal,chatterjee,automaticd,adri,whitelam,aurell}. \GF{We then use this benchmark to evaluate how close conventional strategies can get to the fundamental limit of performance for such systems} and analyze how the presence of realistic \GF{(smoothed)} delta-like functions affects the engine’s performance, providing deeper insights into the optimization of work extraction in nonequilibrium Brownian systems. This study establishes the OS as a benchmark for future research, setting a theoretical upper bound on the performance of collisional Brownian engines and paving the way for further advancements in engine optimization.


This paper is organized as follows: In Section \ref{two}, we present the model, introduce the key analytical expressions for the most general (underdamped) regime, and outline the optimization procedures. Section \ref{three} is devoted to deriving the OS alongside explicit formulas for key thermodynamic quantities, such as power, efficiency, and entropy production. In Section \ref{four}, we benchmark the OS against well-established \GF{strategies} from the literature. Section \ref{five} examines the realistic impact of finite-width delta functions on the performance of the optimal Brownian engine. Finally, conclusions and perspectives are summarized in Section \ref{conc}.

\section{Thermodynamic Overview of Collisional Brownian Engines}
\label{two}

Consider a Brownian engine where a \GF{single} particle, confined in a harmonic potential, interacts with a thermal reservoir at temperature \GF{$T$}. The engine operates in \GF{a cycle of total period $\tau$}, during which a time-dependent external force is applied to inject or extract work from the system. In order to find the OS for this type of engine, we consider the most general case of an underdamped Brownian particle of mass $m$ trapped in a harmonic potential of (fixed) stiffness $k$ and drived by a periodic external force $F(t)$. This system evolves according to

\begin{equation}
    \label{lang}
    m\,\ddot{x}_i = -\gamma\,\dot{x}_i+ \Tilde{F}_i(x,t) + \xi_i(t) 
\end{equation}

\noindent where \GF{we label the strokes by $i \in \{1,2\}$, so that $x_i$ and $\dot{x}_i$ denote the particle's position and velocity during the $i$-th stroke, which evolves under the corresponding reservoir parameters}. Explicitly, $\gamma$ represents the viscous coefficient, $\Tilde{F}_i(x,t) = -k\,x_i +F(t)$ and $\xi_i(t)$ is the stochastic force following the standard white noise properties $\langle \xi_i(t)\rangle = 0$ and $\langle\xi_i(t)\xi_j(t')\rangle = 2\,\gamma\,k_b\,T\,\delta_{ij}\,\delta(t-t')/m$. \GF{This framework serves as a minimal representation for modeling, e.g., celular nanoscale transport processes in biological systems \cite{lopez2008realization,prost1994asymmetric,astumian1994fluctuation}, the stepwise motion of a single kinesin motor along a biopolymer track or a diffusing particle in a ratchet mechanism, which can be realized assigning specific forces $F_i(t)$ to each stroke, such as chemical gradients in active systems or by optical trapping in passive setups \cite{lopez2008realization,Duet}}.

\GF{The probability of finding the particle in a microstate defined by position $x$ and velocity $\dot{x}$ at time $t$ during the $i$-th stroke, \( P_i(x, \dot{x}, t) \), follows the Fokker-Planck-Kramers equation \cite{seifert2012stochastic}}
\begin{equation}
\label{FK}
    \frac{\partial P_i}{\partial t} = -\left[\dot{x}\frac{\partial P_i}{\partial x}+ \Tilde{F}_i(x,t)\frac{\partial P_i}{\partial \dot{x}}+\frac{\partial J_i}{\partial \dot{x}} \right].
\end{equation}

\noindent \GF{Here, $J_i$ is the net probability current, with a drift term $- \gamma \dot{x} P_i$ from viscous friction that slows the particle, producing heat that flows to the bath, and a diffusion term, proportional to the probability density gradient, arising from thermal kicks of bath molecules. Explicitly,}

\begin{equation}
    \label{currentsFK}
    J_i = -\gamma\,\dot{x}\,P_i - \frac{\gamma\,k_B\,T}{m}\,\frac{\partial P_i}{\partial \dot{x}}.
\end{equation}

\noindent This framework lead to a Gaussian distribution for \(P_i(x, \dot{x}, t)\) due to the white noise and the linearity of the deterministic forces. In the absence of driving forces (\(F(t) = 0\)), the system relaxes to thermal equilibrium, described by the Boltzmann distribution. Deviations from thermal equilibrium arise when the system is subjected to temperature gradients and/or time-dependent external forces, driving it toward a nonequilibrium steady state (NESS). In this regime, \GF{the energy balance, representing the first law of thermodynamics, follows from Eq.~(\ref{FK}) together with the definition \( U_i(t) = m \langle \dot{x}_i^2\rangle/2 + k \langle x_i^2\rangle/2 \), yielding two distinct contributions \cite{seifert2012stochastic,landientropy},} namely $\dot{U}_i(t) = -\left[\Dot{W}_i(t) +\Dot{Q}_i(t)\right]$, where
\begin{equation}
\label{work}
    \Dot{W}_i(t) = -F(t)\,\langle \dot{x}_i\rangle(t)
\end{equation}

\noindent represents the power delivered from/to the particle by the external driving during stroke \( i \) and

\begin{equation}
    \Dot{Q}_i(t) = \gamma\left(\langle \dot{x}_i^2 \rangle(t) - \frac{k_B\,T}{m}\right).
\end{equation}

\noindent is the heat \GF{flux from/to} the bath during the same stroke. The driving force is split into two parts, each corresponding to one stroke of the cycle, namely

\begin{equation}
F(t) =
\begin{cases}
    X_1\,g_1(t), & 0 \leq t < \tau_1, \\
    X_2\,g_2(t), & \tau_1 \leq t < \tau,
\end{cases}
\end{equation}

\noindent where $X_i$ denotes the strength of the thermodynamic force applied \GF{during the $i$-th stroke, and $g_i(t)$ defines the temporal profile of the protocol throughout that stroke}. \noindent\GF{Engines in which one form of work is converted into another are commonly referred to as work-to-work converters, systems where part of the work performed by the driving force in one stroke is directly converted into useful work in the subsequent stroke}. This class of systems has been extensively studied in the literature, both theoretically \cite{seifert2012stochastic,pebeu,fisher1999force,busiello2022hyperaccurate,hooyberghs2013efficiency} and experimentally \cite{liepelt2007kinesin,liepelt2009operation,Duet}\GF{, with the main goal being to maximize the conversion of input energy per time (e.g., chemical) into useful output energy per time (e.g., mechanical), as exemplified by cellular transport processes~\cite{lopez2008realization,prost1994asymmetric,astumian1994fluctuation}.} 

When averaged over a cycle, \hbox{\( \overline{\dot{U}} = \overline{\dot{W}}_1 + \overline{\dot{W}}_2 + \overline{\dot{Q}}_1 + \overline{\dot{Q}}_2 = 0 \)}, consistent with the first law of thermodynamics. The second law of thermodynamics is derived by taking the time derivative of the Shannon entropy \( S_i = -k_B \langle \ln P_i \rangle \), along with Eq.~(\ref{FK}), leading to the relation \( \dot{S}_i = \sigma_i(t) - \Phi_i(t) \), where the entropy production $\sigma_i(t)$ and the entropy flux $ \Phi_i(t)$ are given by \cite{seifert2012stochastic,schn}

\begin{equation}
\label{epeq}
   \sigma_i(t) = \frac{m}{\gamma\,T}\int \frac{J_i^2}{P_i}\,dxd\dot{x}\,\,\,\,\,\,\,\,\,\,\textrm{and}\,\,\,\,\,\,\,\,\,\, \Phi_i(t) = \frac{\Dot{Q}_i(t)}{T}.
\end{equation}

\noindent Consistent with the second law, $\sigma_i(t)$ is always non-negative. \GF{Using Eqs.~(\ref{epeq}) in the NESS, where $\sigma_i(t) = \Phi_i(t)$,} one find that the total entropy production rate over a cycle, $\overline\sigma$, is given by
\begin{equation}
\label{eppp}
     \overline{\sigma} =\frac{-(\overline {\dot W}_{1} + \overline {\dot W}_{2})}{T}.
\end{equation}


\noindent In the linear response regime, Eq.~(\ref{eppp}) can be expressed as the sum of thermodynamic forces $f_i = X_i/T$ times thermodynamic fluxes $J_i$, namely $\Bar{\sigma} = J_1\,f_1 + J_2\,f_2$ \cite{seifert2012stochastic,DeGroot1962}, where

\begin{equation}
\label{w11}
\overline {\dot W}_{1} = - TJ_1f_1 = -T(L_{11}\,f_1^2 + L_{12}\,f_1\,f_2),
\end{equation}
\begin{equation}
\label{power}
    \overline {\dot W}_{2} = - TJ_2f_2 = -T(L_{22}\,f_2^2 + L_{21}\,f_2\,f_1),
\end{equation}

\noindent in which $L_{ij}$'s represent the Onsager coefficients. Three points are worth highlighting. \GF{First, entropy production is entirely dictated by the external driving forces: when the driving performs negative work, injecting energy into the system, entropy production increases; conversely, when it performs positive work, with the particle delivering useful output, entropy production decreases}. Second, since we are dealing with work-to-work converters, no coupling exists between  work fluxes and heat flux, implying that the Onsager coefficients that involve temperature are all zero. Third, expressions in the form of Eqs.~(\ref{w11}) and (\ref{power}) are obtained when we average Eq.~(\ref{work}) over a complete cycle.

Finally, a work-to-work engine converts a portion of the input power $\overline {\dot W}_{\rm{in}}$ into useful output power $\overline {\dot W}_{\rm{out}}$, where $\overline {\dot W}_{\rm{in}} < 0$, $\overline {\dot W}_{\rm{out}} \geq 0$ and $|\overline {\dot W}_{\rm{in}}| > |\overline {\dot W}_{\rm{out}}|$. We can thus define the efficiency of the converter as

\begin{equation}
  \eta \equiv -\frac{\overline {\dot W}_{\mathrm{out}}}{\overline {\dot W}_{\mathrm{in}}}.
  \label{efff}
\end{equation}

\noindent In terms of the Onsager coefficient, where we use $i$ for the input and $j$ for the output force, we have

\begin{equation}
\label{effonsager}
    \eta = -\frac{J_j f_j}{J_i f_i}=  -\frac{L_{jj}\,f_j^2 + L_{ji}\,f_i\,f_j}{L_{ii}\,f_i^2 + L_{ij}\,f_j\,f_i}.
\end{equation}

\noindent As a last comment, we highlight that, as stated in Ref.~\cite{proesmans2016}, the optimized value of both power and efficiency can be obtained solely in terms of the Onsager coefficients. Indeed, for a fixed $f_i$, the force $f_j^{\rm{MP}}$ delivering the maximum power is given by $f_j^{\textrm{MP}} = -L_{ji} f_i/2L_{jj}$, with respective maximum power and efficiency given by

\begin{equation}
    \mathcal{P}_{\textrm{MP}} = T\frac{\,L_{ji}^2}{4\,L_{jj}}f_i^2, \qquad
    \eta_{\textrm{MP}} = \frac{L_{ji}^2}{4\,L_{jj}\,L_{ii} - 2\,L_{ji}\,L_{ij}}.
    \label{maxi_f2}
\end{equation}

\noindent Similarly, one finds the force that optimizes efficiency, as well as the optimal efficiency itself, to be

\begin{equation}
  f_{j}^{ME}=\frac{L_{ii} }{L_{ij} }\left(-1+ \sqrt{1-\frac{L_{ji} L_{ij}}{L_{jj} L_{ii}}}\right)f_i,
  \label{eq:x2meta}
\end{equation}
\begin{equation}
  \eta_{ME}=-\frac{L_{ji}}{L_{ij}}+\frac{2L_{ii}L_{jj}}{L_{ij}^2}\left(1-\sqrt{1-\frac{L_{ij} L_{ji}}{L_{ii} L_{jj}}}\right).
  \label{etame}
\end{equation}

\section{Optimal Control strategy}
\label{three}

With the thermodynamics of Brownian engines established, we now shift our attention to the central result of this work. Inspired by Refs. \cite{seifert2007,seifert2008}, we seek the protocol \( F(t) \) that maximizes the \GF{output work rate of the converter, given by}
\begin{equation}
    \overline {\dot W}_{\rm{out}} = -\frac{1}{\tau}\int_{0}^{\tau_1}F(t)\,\langle\dot{x}\rangle\,dt.
    \label{meanwork}
\end{equation}

\noindent By using the Langevin equation~(\ref{lang}) for the averaged quantities, we can express the mean output work rate as
\begin{align}
\overline{\dot{W}}_{\rm{out}} &= -\frac{1}{\tau}\int_{0}^{\tau_1}\left[\gamma\,\langle \dot{x} \rangle + k\,\langle x \rangle + m\,\langle \ddot{x} \rangle\right]\,\langle \dot{x} \rangle\,dt \nonumber\\
&= -\frac{1}{\tau}\left[\int_{0}^{\tau_1}\gamma\,\langle \dot{x} \rangle^2\,dt + \frac{k}{2}\langle x\rangle^2\Bigg|_{0}^{\tau_1} + \frac{m}{2}\langle \dot{x} \rangle^2\Bigg|_{0}^{\tau_1}\right].
\label{optwork}
\end{align}

\begin{figure*}
    \centering
    \includegraphics[scale=1.5]{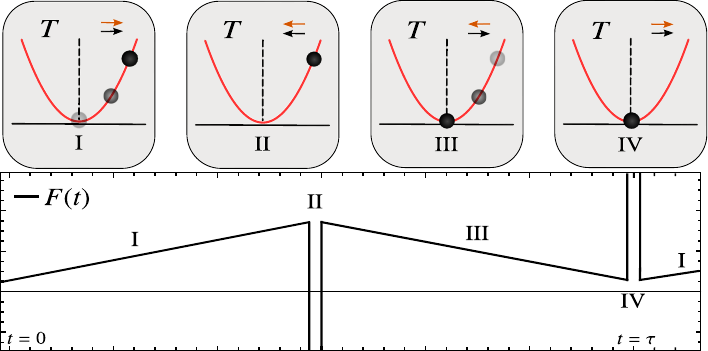}
    \caption{Schematic representation of the optimal Brownian engine (top) and the corresponding optimal protocol \( F(t) \) (bottom). In the top panel, orange and black arrows indicate the direction of the velocity and the external driving force, respectively. The sharp transitions in steps II and IV correspond to delta-like kicks (amplified for better visualization) that instantaneously reverse the velocity, while steps I and III are the main operational stages. Explicitly, during step I, the protocol increases linearly, exactly compensating the natural acceleration to maintain a constant velocity; in step II, a sudden change reverses the velocity. In step III, the protocol decreases linearly to decelerate the particle while keeping its velocity constant; a final sudden change in step IV reverses the velocity again, preparing the system for the next cycle.
 }
    \label{scheme}
\end{figure*}

\noindent 
To determine the protocol, we apply the Euler-Lagrange equation to minimize the remaining integral in Eq.~(\ref{optwork}), where the minus sign in the power expression ensures that minimizing the integrand is equivalent to maximizing the power output. This leads to the condition $\langle \ddot{x}\rangle = 0$, solved by $\langle x_i(t)\rangle = v_i\,t+x_{0i}$, with $v_i = \langle\dot{x}_i\rangle = \rm{constant}$.  Interestingly, the velocity profile that optimizes the performance of the Brownian engine coincides with the one that minimizes the work in a stiffening trap, as shown in previous works \cite{seifert2007,seifert2008}. By applying the boundary conditions $\langle x_1(0)\rangle = \langle x_2(\tau)\rangle$ and $\langle x_1(\tau_1)\rangle = \langle x_2(\tau_1)\rangle$, the number of independent variables is reduced from five, \( \{v_1, v_2, x_{01}, x_{02}, \tau_1\} \), to three, \(\{ v_1, x_{01}, \tau_1\}\). \GF{While the initial position $x_{01}$ is found by maximizing the output work, a subtlety arises when optimizing the stroke duration, $\tau_1$. Maximizing the output work, $\overline{\dot W}_{\rm out}$, with respect to $\tau_1$ leads to the trivial solution $\tau_1 = \tau$. Physically, this occurs because the system extracts the most work when it never has to be ``reset'' for a new cycle, but this does not describe a functional engine. To determine the optimal duration for a useful, cyclic two-stroke engine, a different physical figure of merit is needed. For this reason, we choose to maximize the engine's efficiency, which yields the non-trivial and symmetric result $\tau_1 = \tau/2$ (see Appendix~\ref{derivation} for a detailed derivation of these results).}


\GF{At this point, it is thus crucial to distinguish the two levels of our optimization. First, the shape of the protocol is derived from maximizing work/power, ensuring it is inherently a maximum power protocol. Second, the duration of its strokes is determined by maximizing efficiency, yielding a functional and efficient two-stroke engine cycle. At the same time,} it is important to emphasize that the boundary conditions enforced by the optimization procedure require the velocity to switch from $v_1$ to $v_2 = -v_1$ at $t = \tau/2$, and back again at $t = \tau$ \GF{(see Appendix~\ref{derivation})}. Since the velocities must remain constant during each stroke, maintaining this condition requires impulsive forces at the transitions, i.e.,

\begin{figure}
    \centering
    \includegraphics[width=1\linewidth]{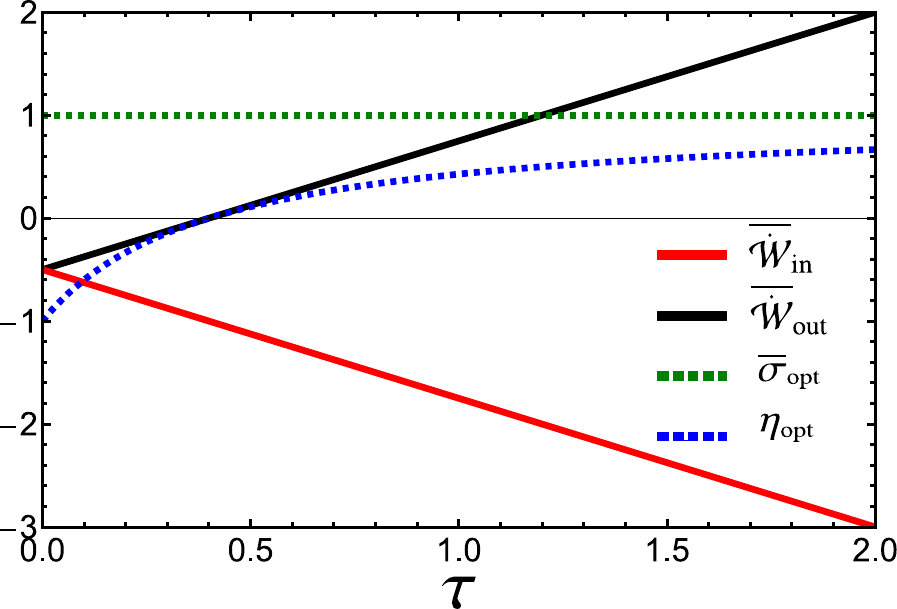}
    \caption{Input work (red solid line), output work (black solid line), efficiency (blue dashed line), and entropy production (green dashed line) of the optimal \GF{strategy} as functions of the driving period $\tau$. The parameters were set to $\gamma = 1$, $T = 1$, $v_1 = 1$ and $k = 10$ to enhance visual clarity.}
    \label{performance}
\end{figure}

\begin{equation}\label{deltaforce}
    F(t) \propto 2\,m\,v_1\left(\delta(t-\tau)-\delta(t-\tau/2)\right).
\end{equation} 

\noindent Note that the presence of inertia is essential for the cycle to operate properly, as the velocity jumps between strokes require impulsive forces proportional to the system’s mass. This implies that the OS can only be realized in systems where inertial effects play a significant role. With all constants determined, the OS is found to be $F(t) = F_{\rm{out}}(t) + F_{\rm{in}}(t)$, where $F_{\rm{out}}$ and $F_{\rm{in}}$ act on $0 \leq t \leq \tau/2$ and $\tau/2 \leq t \leq \tau$, respectively, and are given by

\begin{equation}\label{f1}
    F_{\rm{out}}(t) = \gamma v_1 + kv_1 \left(t-\frac{\tau}{2}\right) -mv_1\left(\delta(t-\tau/2)-\delta(t)\right),
\end{equation}
\begin{equation}\label{f2}
    F_{\rm{in}}(t) =-\gamma v_1 + kv_1\left(\frac{\tau}{2}-t\right) -mv_1\left(\delta(t-\tau/2) - \delta(t-\tau)\right).
\end{equation}

The protocol shape, as well as the system dynamics, can be seen in Fig.~\ref{scheme}. Notably, the external force acts on the system in such a way that the acceleration is zero, exactly canceling the inertial term. However, at the boundaries, the system can only reset by making jumps in velocity, which highlights the essential role of inertia during stroke transitions. Since the inertial contributions are sharply localized (delta-like), they can be neglected in the continuous dynamics, eliminating the dependence of the optimal thermodynamic quantities on mass and allowing the protocol to inject energy into the system at no extra energetic cost. \GF{Indeed, the work performed by these impulsive forces is strictly zero. This is a direct consequence of the optimal protocol's boundary conditions, 
which enforce a velocity reversal ($\pm v_{1} \leftrightarrow \mp v_{1}$). This reversal ensures that the kinetic energy is conserved ($\Delta K = 0$). By the work--energy theorem, the work done during this idealized process (which is analogous to a perfectly elastic collision against a rigid wall) must be zero, despite the infinite nature of the force}. These peaks correspond to two (adiabatic) preparatory steps (II and IV in Fig.~\ref{scheme}) that adjust the system’s velocity before the main strokes (III and I). \GF{This 4-step structure is a direct consequence of the optimization and contrasts with previous works, such as \cite{filho1,mamede2022,mamede,harunariasy,forao1}, where simpler 2-stroke cycles based on non-optimal strategies were studied}. \GF{We emphasize that the assumption of an infinitesimal ``zero duration" delta peak is an idealization imposed by the OS; the substantial energetic cost of a realistic, finite-duration pulse is analyzed in detail in Section~\ref{five}, where we introduce the concept of preparation work.}

By substituting $F(t)$ into Eqs.~(\ref{meanwork}), (\ref{efff}) and (\ref{eppp}), we obtain the expressions for the optimal input and output powers, efficiency, and entropy production as
\begin{equation}
    \label{optworkin}  
   \overline{\dot{\mathcal{W}}}_{\rm{in}} = \frac{-v_1^2\,(k\,\tau+4\,\gamma)}{8}\,\,\,\,\,\,\,\,\textrm{and}\,\,\,\,\,\,\,\,\overline{\dot{\mathcal{W}}}_{\rm{out}} = \frac{v_1^2\,(k\,\tau -4\,\gamma)}{8}
\end{equation}
\begin{equation}
    \label{opteff}
   \eta_{\rm{opt}} = 1-\frac{8 \gamma }{4 \gamma +k  \tau}
\end{equation}
\begin{equation}
    \label{optep}
    \overline{\sigma}_{\rm{opt}} = \frac{\gamma}{T}\,v_1^2
\end{equation}

\begin{figure*}
    \centering
    \includegraphics[scale=0.372]{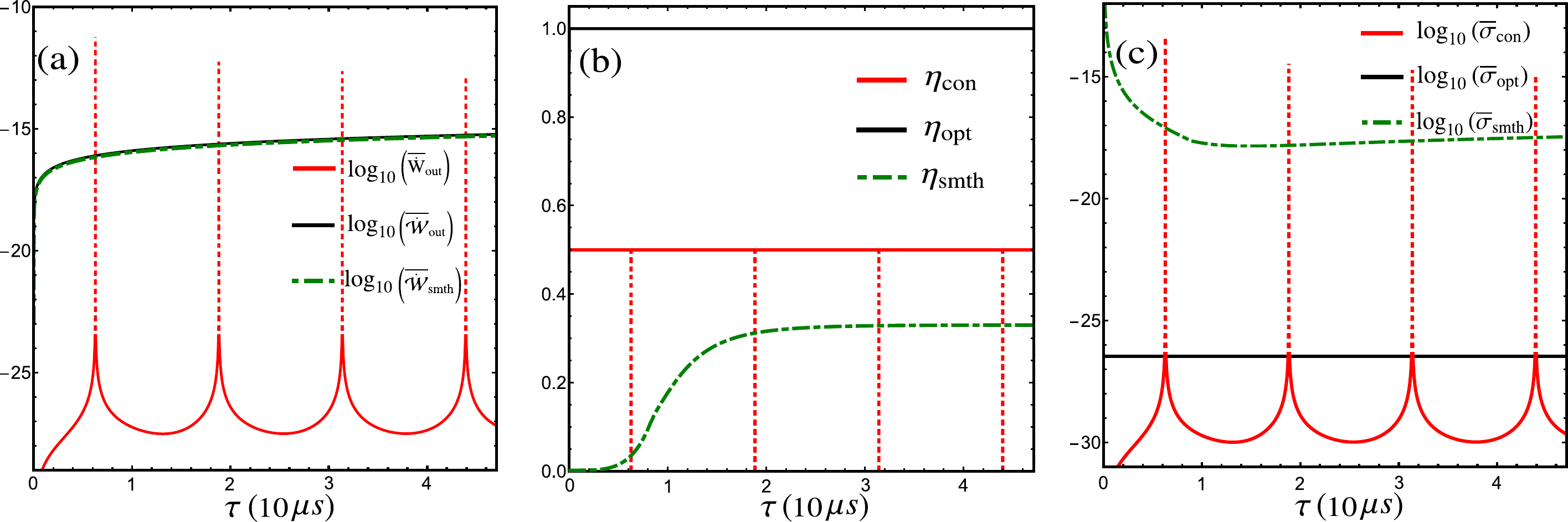}
    \caption{Panels (a), (b), and (c) show, respectively, the base-10 logarithm of the power, the efficiency, and the base-10 logarithm of the entropy production as functions of the driving period, for both the constant (red lines), optimal (black lines) and smoothed (green lines) protocols. The latter was computed using $d\tau = 0.01\,\tau$, and its results will be further discussed in Sec.~\ref{five}. Resonance intervals are indicated by red dashed regions. All quantities were computed using the experimental parameters mentioned in the main text. Logarithmic scaling was used in panels (a) and (c) to enhance visualization and allow for a clearer comparison between the \GF{strategies}.}
    \label{compar}
\end{figure*}

Fig.~\ref{performance} highlights fundamental characteristics of the system’s optimal performance. First, the existence of an external potential is crucial for the system to function as an engine, since the condition $\eta_{\rm{opt}} > 0$ requires $k\,\tau > 4\gamma$. Such a requirement is typically fulfilled in experimental underdamped systems, where $k\tau \gg \gamma$. Second, the output work is always smaller than the absolute input work, i.e., $\overline{\dot{\mathcal{W}}}_{\rm{out}} < |\overline{\dot{\mathcal{W}}}_{\rm{in}}|$, meaning that no regime change is observed in the optimal limit, with the system consistently assigning the first stroke as the useful one and the second as the charging one — a feature that notably differs from previous works where transitions between operational regimes could emerge \cite{forao1}. Third, since the input and output work scale as \(-\kappa\tau\) and \(+\kappa\tau\), respectively, this contribution vanishes in the expression for the entropy production, leaving it solely dependent on \(\gamma\). As a result, power and efficiency can be enhanced by tuning $k$ and/or $\tau$ without increasing the entropy production. Fourth, the boundary conditions constrain the dynamics such that only a single thermodynamic force remains (the one proportional to the impulsive force responsible for changing the system's velocity) which results in the merging of the Onsager coefficients, leading to the expressions in the form $\overline{\dot{\mathcal{\sigma}}} = (L_{ii} + L_{ij} + L_{ji} + L_{jj})f^2 \GF{\equiv L_{\sigma}\,f^2}$ for the entropy production and $\overline{\dot{\mathcal {W}}}_i = (L_{ii} + L_{ij})f^2 \GF{\equiv L_{Wi}\,f^2} $ for work, where $f \propto 2\,m\,v_1\,\delta(t-t')$ and can be determined analytically once the duration of the preparatory steps is specified.  

Finally, since the entropy production in each stroke satisfies \(\overline{\sigma}_i \propto -\overline{\dot{W}}_i\), maximizing (minimizing) the average work naturally leads to a minimization (maximization) in entropy production. Indeed, in the output stroke, the optimal force is designed to extract the maximum possible work from the system, which leads to a global minimum in entropy production. In the input stroke, the goal is to restore the system using the least amount of energy, which means making the work input as small (in magnitude) as possible. This results in a local minimum in entropy production, constrained by the boundary condition $v_2 = -v_1$. Because of this constraint \GF{(imposed by the OS to ensure that the system operates as a functional two-stroke engine)}, the total entropy production is not globally minimized across all possible protocols, but it is minimized within the class of protocols that respect this symmetry. In fact, \(\overline{\sigma}_i\) typically scales with \(|\overline{W}_i|\), so protocols that produce less work often dissipate less than $\overline{\sigma}_{\rm{opt}}$. The advantage of the OS is that it can produce much more work at a low dissipation than general protocols. 

\section{Comparison with other strategies}
\label{four}
\GF{This section aims to use the OS derived in Sec.~\ref{three} as an upper bound to quantify the performance gap between the theoretical limit and (realistic) conventional strategies, which have been the subject of extensive recent study}. Although \GF{protocols} in the overdamped regime have been more widely investigated \cite{filho1,mamede2022,harunariasy,angel,Noa1}, it has recently been shown that underdamped systems can exhibit resonant effects that markedly boost performance \cite{forao1}, achieving power and efficiency beyond those attainable in the overdamped regime. Specifically, resonances occur when

\begin{equation}
    \tau_{\mathrm{res}} = \frac{2\,\pi\,n}{\sqrt{k/m}}, \quad n =1,\,2,\,3...
\end{equation}

\noindent Within this context, the most commonly studied drivings are (time) constant, linear, and periodic protocols. Among these, \GF{previous works \cite{forao1} suggest that} the constant protocol \GF{(in the underdamped regime)} stands out for consistently delivering the best quantitative results, which is why we adopt it as our primary benchmark. Linear and periodic drivings lead to qualitatively similar behaviors and are therefore discussed only briefly. We also include a few remarks on the performance of overdamped systems for comparison. The main analytical expressions for the protocols, as well as for both the underdamped and overdamped regimes, are provided in Appendix~\ref{appendixunder} and Appendix~\ref{appendixover}, respectively. To ensure a fair and physically grounded comparison, all performance indicators are evaluated at maximum power, as defined in Eq.~(\ref{maxi_f2}). For the numerical values, we choose experimentally realistic parameters inspired by previous studies on Brownian engines \cite{martinez2016brownian,Duet}: $T = 300\rm{\,K}$, $\gamma \approx 10^{-20}\text{s}^{-1}$, $k \approx \text{pN/$\mu$m}$, $m \approx 10^{-18}\text{kg}$, $\tau \approx 1\text{$\mu$s}$ and $f_i \approx 1\,\text{fN}$, which correspond to typical velocities around $v \approx 1\,\text{cm/s}$.

Insights into the distinct thermodynamic behaviors of the optimal and constant protocols can be drawn from Fig.~\ref{compar}. In panel (a), the OS exhibits a monotonic increase in output power with period, maintaining robust performance around \(10^{-16}\,\mathrm{J/s}\). In contrast, the constant protocol outperforms the OS only within narrow resonance windows (dashed areas), peaking near \(10^{-11}\,\mathrm{J/s}\) and dropping sharply in power outside these regions. Panel (b) shows that the OS sustains near-unity efficiency \GF{($\eta = 1-8\gamma/(4\gamma+\kappa\tau) \approx 1-10^{-8} = 0.99999999$)} across all periods—a striking result, especially considering that this value is achieved under conditions of global maximum power. The constant protocol, in turn, holds an efficiency near 0.5 for most periods, exhibiting abrupt transitions into non-converting regimes near resonance (which manifest as vertical lines in panel (b)) followed by a return to \(\eta = 0.5\) at exact resonance. In panel (c), the OS maintains a constant entropy production of about \(10^{-26}\,\mathrm{J/Ks}\), roughly ten orders below its power output, reflecting a minimal and controlled dissipation. The constant protocol shows low dissipation most of the time but develops sharp entropy peaks near resonance, reaching up to \(10^{-13}\,\mathrm{J/Ks}\), indicating that its high performance under resonance comes at a significant thermodynamic cost.

\begin{figure}
     \centering
    \includegraphics[width=\linewidth]{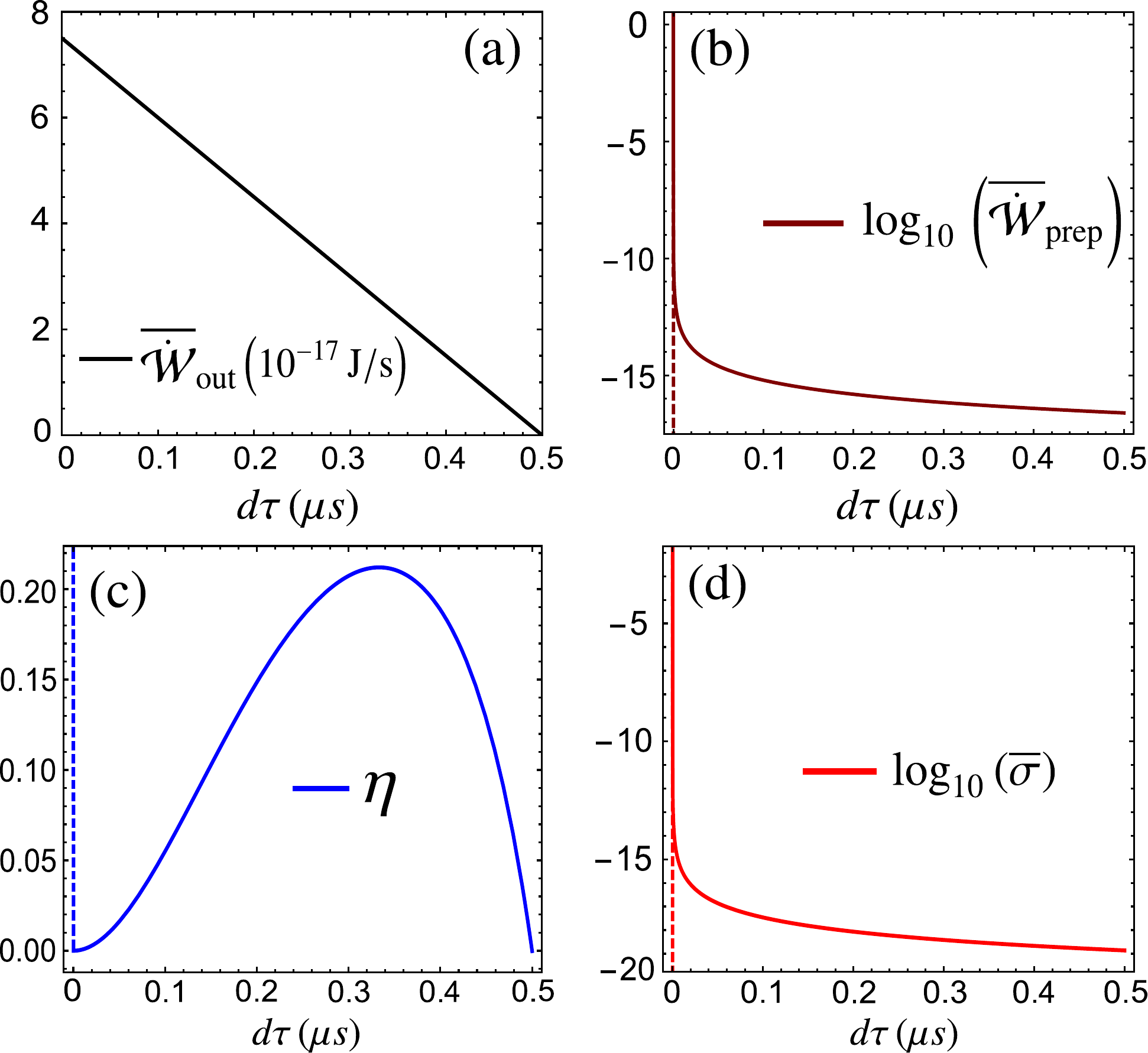}
    \caption{Performance of the engine as a function of the regularization width $d\tau$ of the delta contributions. (a) Output power $\overline{\mathcal{\dot{W}}}_{\rm{out}}$; (b) Log-10 of the preparation work $\overline{\mathcal{\dot{W}}}_{\rm{prep}}$; (c) efficiency $\eta$; (d) Log-10 of the entropy production $\overline{\sigma}$. The dashed regions in panels (b), (c), and (d) indicate the discontinuity in thermodynamic quantities occurring at \( d\tau = 0 \). We adopt the same parameters as in Fig.~\ref{compar}, fixing the driving period at \( \tau = 6\,\mu\text{s} \).}
    \label{delta}
\end{figure}

For completeness, we briefly compare the performance of other \GF{strategies} in both damping regimes. In the underdamped regime, the periodic protocol stands out with \hbox{\( W_{\text{out}} \approx 5.3 \times 10^{-12} \,\text{J/s} \)} and \( \eta \approx 0.51 \), albeit with an entropy production \( \overline{\sigma} \approx 1.7 \times 10^{-14} \,\text{J/Ks} \), close to that of the constant protocol. The linear protocol performs modestly, with $W_{\text{out}} \approx 1.75\times10^{-23}$~J/s, $\overline{\sigma} \approx 5.8\times10^{-26}$~J/Ks, and $\eta \approx 0.5$. In the overdamped case ($\gamma \approx 10^{-8}\,\rm{s^{-1}}$), the constant protocol exhibits the best performance, yielding $W_{\text{out}} \approx 6\times10^{-24}$~J/s, $\overline{\sigma} \approx 2\times10^{-26}$~J/Ks, and efficiency $\eta \approx 0.5$. The linear and periodic drivings, in contrast, deliver significantly lower performance, with output powers several orders of magnitude below those of the constant protocol.

\GF{In summary, the comparative analysis in this section reveals key physical insights. In the underdamped regime, standard strategies can achieve high power by exploiting the system’s mechanical resonances; however, this approach is thermodynamically costly, leading to large spikes in entropy production and unstable efficiency near resonance. Overdamped engines, on the other hand, remain far from optimal since, in this regime, the external potential exerts only a negligible influence on the dynamics, which reduces the requirement for the driving force to counteract it and, consequently, results in less work production. In contrast, the OS demonstrates that operating with an external potential while avoiding the resonant regime allows symmetries that constitute the most effective \GF{strategy} for achieving optimal performance. Indeed, the protocol’s design enforces constant velocity satisfying \(v_1 \leftrightarrow -v_1\) and equal-duration strokes, maximizing power output and imposing a symmetry between \(\overline{\mathcal{\dot{W}}}_{\rm{in}}\) and \(\overline{\mathcal{\dot{W}}}_{\rm{out}}\) that differs solely by the viscous dissipation term \(4\gamma\) (see Eq.~(\ref{optworkin})). This structural symmetry substantially enhances power and efficiency while reducing entropy production.}


\section{Impact of realistic Delta-Like Forces}
\label{five}

\GF{Having established the theoretical performance benchmark in the ideal limit of infinitesimal impulses,} we now replace the deltas in \( F(t) \) with narrow Gaussians, \( \delta(t) \to \mathcal{N}(t,\,d\tau) \), with variance \GF{(width)} $d\tau$ and centered at the same switching times \GF{$t$}, \GF{better reflecting realistic, smooth force exchanges.} \GF{Note that, with this substitution, the ideal limit of instantaneous delta-like impulses discussed in Sec.~\ref{three} is recovered only when $d\tau = 0$}. Although this substitution makes the mean Langevin equation less tractable analytically, in the experimental regime of interest (\( \gamma \ll k \), \( m \ll k \)), the approximation \( x(t) \approx F(t)/k \) is accurate. This yields \( v(t) \approx \dot{F}(t)/k \), and the optimal work per stroke becomes

\begin{equation}
   \overline{\mathcal{\dot{W}}}_i \approx -\frac{1}{\tau} \int_{\tau_a}^{\tau_b} F(t)\,\frac{\dot{F}(t)}{k}\,dt = -\frac{1}{2k\tau}F(t)^2\biggr|_{\tau_a}^{\tau_b}.
\end{equation}

The results obtained using this approximation show excellent agreement with numerical results. Here, it is important to notice that the expressions for work have now four contributions, two of them identical to Eq.~(\ref{optworkin}) and two accounting for the preparatory steps (Gaussian contributions) arising from the loss of adiabaticity. The latter is not considered useful work, but rather an energetic cost required to sustain the cycle, since it occurs over a very short time interval and no extractable work can be obtained from it. This leads to an additional term, $\overline{\mathcal{\dot{W}}}_{\rm{prep}}$, modifying Eqs.~(\ref{eppp}) and (\ref{efff}) as

\begin{equation}
    \overline{\sigma} = \frac{-\left(\overline{\mathcal{\dot{W}}}_{\rm{in}}+\overline{\mathcal{\dot{W}}}_{\rm{out}}+\overline{\mathcal{\dot{W}}}_{\rm{prep}}\right)}{T}
\end{equation}
\begin{equation}
    \eta = \frac{\overline{\mathcal{\dot{W}}}_{\rm{out}}}{\left|\overline{\mathcal{\dot{W}}}_{\rm{in}}\right| + \left|\overline{\mathcal{\dot{W}}}_{\rm{prep}}\right|}
\end{equation}

\GF{Figure~\ref{delta} demonstrates the influence of delta-like pulses on the engine’s performance, revealing a fundamental trade-off between the OS and the energetic cost of implementing it. The output power (panel (a)) initially matches the optimal theoretical value but decreases nearly linearly as the pulse duration \( d\tau \) increases, ultimately vanishing around \( d\tau = \tau/12 \), which suggests a physical lower bound on how long abrupt forces can act. This decline occurs because any finite pulse duration breaks the instantaneous velocity reversal required by the OS (\(v_1 \leftrightarrow -v_1\)), causing deviations from the optimal regime that reduce the extractable power. The preparation work \( \overline{\mathcal{\dot{W}}}_{\rm{prep}} \) (panel (b)), in turn, is considerably high for very short pulses due to the fact that reversing the particle’s velocity demands a large impulsive force applied over a very brief interval \( d\tau \) (\GF{$F_{\rm{imp}} \propto 1/d\tau$}), making the work cost diverge as \( d\tau \to 0 \). As the pulse duration grows, the force magnitude and corresponding energetic cost decrease sharply. The efficiency \( \eta \) (panel (c)) reflects these competing effects, rising from near zero at very small \( d\tau \), where preparation costs dominate, to a peak at an intermediate \( d\tau \) value that balances power output and energetic expense, before decreasing again as power output continues to decline. Entropy production \( \overline{\sigma} \) (panel (d)) closely follows the behavior of the preparation work, indicating that dissipation is largely governed by the energetic demands of the \(v_1 \leftrightarrow -v_1\) condition.}

Importantly, the dashed regions denote abrupt transitions at exactly \( d\tau = 0 \), \GF{where $\overline{\mathcal{\dot{W}}}_{\rm{prep}}$ is strictly zero (since $\mathcal{N}(t,0) = 0$) and both entropy production and efficiency sharply reach their theoretical optima} (entropy production around \(10^{-26} \mathrm{J/Ks}\) and efficiency \(\eta = 1-10^{-8} = 0.99999999\)), \GF{confirming the values of the ideal limit obtained in Sec.~\ref{four}}. This demonstrates that truly optimal performance requires infinitely sharp, effectively instantaneous pulses, \GF{an idealization not achievable in practice}.

Overall, these results indicate that, although smoothing the delta-like kicks makes the implementation of the OS experimentally feasible, the intense impulsive forces required to prepare the system cause the preparation work to diverge as \( d\tau \to 0 \), driving the efficiency toward zero and sharply increasing entropy production. When compared with resonant engines, the OS is less efficient but delivers higher and controllable power, even under realistic conditions (\(d\tau > 0.05\,\mu s\)), making it an effective and reliable driving strategy. 
In fact, as shown in Fig.~\ref{compar}, for a fixed \( d\tau = \tau/100 \), the smoothed protocol attains power values very close to the optimal one as \(\tau\) varies, although with a lower efficiency (approximately 0.33) and higher dissipation (around \(10^{-17}\,\mathrm{J/Ks}\)).


\section{Conclusions}
\label{conc}

In this work, we derived the \GF{optimal control strategy} for a collisional Brownian engine driven by external forces. By variationally maximizing the output work, we showed that the OS consists of piecewise-linear force segments that maintain constant velocity by canceling the particle’s acceleration, separated by impulsive contributions that instantaneously reverse its direction of motion. This structure globally maximizes the power output while locally minimizing entropy production. Analytical expressions for key thermodynamic quantities, such as output power, efficiency, and entropy production were obtained, allowing for a complete characterization of the engine’s performance. The resulting protocol displays exceptional thermodynamic behavior: it achieves both maximum work and near-optimal efficiency, with entropy production remaining several orders of magnitude smaller than the output power. When compared to standard drivings, the OS outperforms them globally in both efficiency and power, while being locally surpassed in power only near resonance. Moreover, its entropy production remains significantly lower than the magnitude of power output, setting it apart from the alternative \GF{strategies}. To investigate experimental feasibility, we analyzed a more realistic scenario in which the impulsive contributions are smoothed into narrow Gaussian pulses, corresponding to a finite but small time interval $d\tau$. Although performance degrades for non-infinitesimal $d\tau$, with efficiency decreasing and entropy production increasing compared to the ideal delta case, the protocol still achieves high power robustly and maintains entropy production at manageable levels, demonstrating its viability for practical implementations.

Altogether, our findings establish a new benchmark for finite-time thermodynamic optimization of collisional Brownian engines, highlighting the importance of carefully engineered drivings to optimize the engine performance. They also provide practical insights for designing microscale engines capable of combining high power and low dissipation under realistic physical constraints.

\acknowledgments 

\noindent Authors gratefully acknowledge Yasmin Torres for useful hints in Fig.~\ref{scheme} and Fernando S. Filho for a detailed reading of the manuscript. The financial support from FAPESP under grant number 2022/16192-5 is also acknowledged.
\appendix
\section{Derivation of the Optimal Control Strategy}\label{derivation}

\GF{We begin by applying Euler-Lagrange in Eq.~(\ref{optwork}). This gives us}
\GF{\begin{align}
    x_1(t) &= v_1 t + x_{01}, \label{eq:oldeqapp1} \\
    x_2(t) &= v_2 t + x_{02},\label{eq:oldeqapp2}
\end{align}}
\noindent\GF{where $v_1$, $x_{01}$, $v_2$, and $x_{02}$ are constants to be determined.  
The boundary conditions}
\GF{\begin{align}
    x_1(0) &= x_2(\tau), \\
    x_1(\tau_1) &= x_2(\tau_1),
\end{align}}
\noindent\GF{lead to}
\GF{\begin{align}
    v_2 &= \frac{v_1\,\tau_1}{\tau_1 - \tau},\label{eq:neweqapp1} \\
    x_{02} &= \frac{x_{01}\,(\tau -\tau_1) + v_1\,\tau\,\tau_1}{\tau - \tau_1}.\label{eq:neweqapp2}
\end{align}}
\noindent\GF{reducing the problem from five to three independent parameters $(v_1, x_{01}, \tau_1)$. Plugging Eqs.~(\ref{eq:neweqapp1}),(\ref{eq:neweqapp2}) in Eq.~(\ref{eq:oldeqapp2}) and using Eq.~(\ref{optwork}) of the main text (with proper integration limits), one arrives at expressions for the first and second strokes, respectively:}
\GF{\begin{align}
    \overline{\dot{W}}_{\rm{out}} &= \frac{\tau_1}{2\tau} \left(2\,v_1^2\,\gamma\,+\,v_1\,k\,(2x_{01}\,+\,v_1\,\tau_1) \right), \\
    \overline{\dot{W}}_{\rm{in}} &= \frac{\tau_1}{2\tau} \left( 2v_1\,x_{01}\,k +\,v_1^2\,k\,\tau_1\,-\,\frac{2v_1^2\,\gamma\,\tau_1}{\tau\,-\,\tau_1} \right).
\end{align}}

\noindent\GF{Maximizing $\overline{\dot{W}}_{\rm out}$ with respect to $v_1$ (corresponding to the physically admissible branch of the optimization problem) yields $x_{01} = -v_1 \tau_1$. Substituting,}
\GF{\begin{align}
    \overline{\dot{W}}_{\rm{out}} &= \frac{v_1^2 \tau_1}{2\tau} \left( -2\gamma + k \tau_1 \right),\label{eq:neweqapp3} \\
    \overline{\dot{W}}_{\rm{in}} &= -\frac{v_1^2 \tau_1^2}{2\tau} \left(k + \frac{2\gamma}{\tau - \tau_1} \right)\label{eq:neweqapp4}.
\end{align}}

\noindent\GF{From the above expressions, it is clear that $\overline{\dot{W}}_{\mathrm{out}}>0$ when $\kappa\tau_1 > 2\gamma$, while $\overline{\dot{W}}_{\mathrm{in}}<0$, unambiguously identifying the first stroke as the useful one and the second as the loading one. As shown in the main text, the system functions as an engine only in the underdamped regime, where $\kappa\tau_1 > 2\gamma$ (and, in fact, $\kappa\tau_1 \gg 2\gamma$ in such systems (see Sec.~\ref{four}), ensuring this condition is met).}

\GF{Maximizing $\overline{\dot{W}}_{\rm{out}}$ (minimizing $\overline{\dot{W}}_{\rm{in}}$) with respect to $\tau_1$ yields only the solutions $\tau_1 = 0$ or $\tau_1 = \tau$, corresponding to the (trivial) single-stroke case discussed in the main text. To obtain a functional two-stroke engine, we instead maximize the efficiency $\eta = -\overline{\dot{W}}_{\rm{out}}/\overline{\dot{W}}_{\rm{in}}$ (with $\overline{\dot{W}}_{\rm{out}}$ and $\overline{\dot{W}}_{\rm{in}}$ given by Eqs.~(\ref{eq:neweqapp3}) and (\ref{eq:neweqapp4}), respectively), which gives, already considering the underdamped limit,}
\GF{\begin{equation}
    \tau_1 = \frac{\tau}{2}.
\end{equation}}

\GF{Eqs.~(\ref{eq:oldeqapp1}) and (\ref{eq:oldeqapp2}) thus becomes}
\GF{\begin{align}
    x_1(t) &= v_1 \left( t - \frac{\tau}{2} \right), \\
    x_2(t) &= v_1 \left( \frac{\tau}{2} - t \right),
\end{align}}

\noindent\GF{implying velocities jumping from $\pm v_1$ to $v_2 = \mp v_1$ at $t = \tau/2$ and $t = \tau$ ($t = 0$), respectively. These jumps justify the need for impulsive delta-like forces, given by Eq.~(\ref{deltaforce}). Substituting these expressions for position and velocity, together with Eq.~(\ref{deltaforce}), into the averaged Langevin equation yields the driving forces in Eqs.~(\ref{f1})-(\ref{f2}) of the main text. Using these forces in Eqs.~(\ref{meanwork}), (\ref{efff}), and (\ref{eppp}) provides the thermodynamic quantities in Eqs.~(\ref{optworkin})–(\ref{optep}).}
\newline
\section{Main expressions for general underdamped Brownian engines}
\label{appendixunder}

Building on the framework developed in Ref.~\cite{forao1}, the relevant thermodynamic expressions retain the same structure as those derived in Sec.~\ref{two}. A general time-dependent driving can be represented as $\mathcal{F}(t) = f_i\,g_i(t)$, where

\begin{equation}
    g_1(t) = \frac{a_0}{2}+\sum\,a_n\,\cos\left({\frac{2\pi\,n\,t}{\tau}}\right)+b_n\,\sin\left({\frac{2\pi\,n\,t}{\tau}}\right)\,\,\,\,\rm{and}
\end{equation}

\begin{equation}
    g_2(t) = \frac{c_0}{2}+\sum\,c_n\,\cos\left({\frac{2\pi\,n\,t}{\tau}}\right)+d_n\,\sin\left({\frac{2\pi\,n\,t}{\tau}}\right)
\end{equation}

\noindent The driving protocol is defined by selecting the Fourier coefficients \( a_n, b_n, c_n, \) and \( d_n \), such that \( c_n = d_n = 0 \) for the first step (\( i = 1 \)) and \( a_n = b_n = 0 \) for the second step (\( i = 2 \)). Notably, the Fourier expansion inherently satisfies the required boundary conditions, i.e., the continuity of the probability distribution at \( t = \tau/2 \) and the periodicity condition ensuring that the system returns to its initial state at \( t = \tau \). Averaging Eq.~(\ref{lang}) yields the following general expression for the mean velocity \( \bar{v}(t) \):

\begin{equation}
\begin{split}
\bar{v}(t) = \sum_{k=1}^{\infty} &\left( f_1 \cdot a_{1vk} + f_2 \cdot a_{2vk} \right) \cos\left( \frac{2\pi\,k\,t}{\tau} \right) \\
&+ \left( f_1 \cdot b_{1vk} + f_2 \cdot b_{2vk} \right) \sin\left( \frac{2\pi \,k\,t}{\tau} \right)
\end{split}
\label{fourier}
\end{equation}

\noindent where \( a_{ivk} \) and \( b_{ivk} \) denote the Fourier coefficients of the mean velocity, determined by the specific form of the driving forces. By averaging Eq.~(\ref{work}) with \( v(t) \) given by Eq.~(\ref{fourier}), one obtains expressions analogous to Eqs.~(\ref{w11}) and~(\ref{power}). The general form of the Fourier coefficients of the velocity, along with the corresponding Onsager coefficients, is provided by:

$$a_{1kv} = \frac{8\,\pi\, k\,\tau\, \left(8\,\pi \,\gamma \,k\,\tau\, a_n+b_n\,\left(\tau ^2\, (\gamma^2 -\omega_{\textrm{D}}^{2} )-16\,\pi ^2\,k^2\right)\right)}{\left(\tau ^2\,(\gamma -\omega_{\textrm{D}})^2+16\,\pi ^2\, k^2\right) \left(\tau ^2\,(\gamma +\omega_{\textrm{D}} )^2+16\,\pi ^2\,k^2\right)},$$

$$a_{2kv} = \frac{8\,\pi\, k\,\tau\, \left(8\,\pi\, \gamma\, k\,\tau \,c_n\,+\,d_n\,\left(\tau ^2\,(\gamma^2 -\omega_{\textrm{D}}^{2} )-16\,\pi ^2\,k^2\right)\right)}{\left(\tau ^2\,(\gamma -\omega_{\textrm{D}} )^2+16\,\pi ^2 k^2\right)\,\left(\tau ^2\,(\gamma +\omega_{\textrm{D}} )^2+16\,\pi^2\,k^2\right)},$$

$$b_{1kv} = \frac{8\,\pi\, k\,\tau\, \left(a_n\,\left(\tau ^2\,\left(\omega_{\textrm{D}}^{2}-\gamma ^2\right)+16\,\pi ^2\,k^2\right)+8\,\pi\, \gamma\, k\,\tau \,b_n\right)}{\left(\tau ^2\,(\gamma -\omega_{\textrm{D}} )^2+16\,\pi ^2\,k^2\right)\,\left(\tau ^2\,(\gamma +\omega_{\textrm{D}} )^2+16\,\pi ^2\,k^2\right)},$$

$$b_{2kv} = \frac{8\,\pi \,k\,\tau\, \left(c_n\,\left(\tau ^2\,\left(\omega_{\textrm{D}}^i-\gamma ^2\right)+16\,\pi ^2\,k^2\right)+8\,\pi\, \gamma\, k\,\tau \,d_n\right)}{\left(\tau ^2\,(\gamma -\omega_{\textrm{D}} )^2+16\,\pi ^2\,k^2\right)\,\left(\tau ^2\,(\gamma +\omega_{\textrm{D}} )^2+16\,\pi ^2\,k^2\right)},$$
\newline
$$L_{11} = T\sum_{n = 1}^{\infty}\sum_{k=1}^{\infty}\,\frac{\left(\pi\, k\,a_{1kv}\,a_k-b_{1kv}\,\left(a_0\,\left((-1)^k-1\right)-\pi\, k\,b_k\right)\right)}{4\,\pi\, k} +$$ $$(1-\delta_{n,k})\cdot \left(-\frac{\left((-1)^{k+n}-1\right)\,(k\,a_n\,b_{1kv}-n\,a_{1kv}\,b_n)}{2\,\pi\, \left(k^2-n^2\right)}\right),$$

$$L_{12} = T\sum_{n = 1}^{\infty}\sum_{k=1}^{\infty}\frac{\left(\pi\, ka_{2kv}\,a_k-b_{2kv}\,\left(a_0\,\left((-1)^k-1\right)-\pi\, k\,b_k\right)\right)}{4\,\pi\, k}+$$ $$(1-\delta_{n,k})\cdot\,\left(-\frac{\left((-1)^{k+n}-1\right)\,(k\,a_n\,b_{2kv}-n\,a_{2kv}\,b_n)}{2\,\pi\, \left(k^2-n^2\right)}\right),$$

$$L_{21} = T\sum_{n = 1}^{\infty}\sum_{k=1}^{\infty}\frac{1}{4}\,\left(a_{1kv}\,c_k+\frac{c_0\,\left((-1)^k-1\right)\,b_{1kv}}{\pi\, k}+b_{1kv}\,d_k\right) + $$ $$(1-\delta_{n,k})\cdot\,\left(\frac{\left((-1)^{k+n}-1\right)\,(k\,b_{1kv}\,c_n-n\,a_{1kv}\,d_n)}{2\,\pi\, \left(k^2-n^2\right)}\right),$$
and
$$L_{22} = T\sum_{n = 1}^{\infty}\sum_{k=1}^{\infty}\frac{1}{4}\,\left(a_{2kv}\,c_k+\frac{c_0\,\left((-1)^k-1\right)\,b_{2kv}}{\pi\, k}+b_{2kv}\,d_k\right) +$$ $$(1-\delta_{n,k})\cdot\,\left(\frac{\left((-1)^{k+n}-1\right)\,(k\,b_{2kv}\,c_n-n\,a_{2kv}\,d_n)}{2\,\pi\, \left(k^2-n^2\right)}\right),$$

\noindent where $\omega_{\textrm{D}} = \sqrt{\gamma^2 - 4\,\kappa}$ is the damped oscillation frequency of the system. The drivings considered include constant, linear, and periodic protocols, which are respectively defined as follows:

\[
g_i(t) =
\begin{cases}
1, & 0 < t \leq \tau/2, \\
-1, & \tau/2 < t \leq \tau.
\end{cases}
\]

\[
g_i(t) =
\begin{cases}
\lambda t, & 0 < t \leq \tau/2, \\
\lambda (t - \tau/2), & \tau/2 < t \leq \tau.
\end{cases}
\]

\[
g_i(t) =
\begin{cases}
\cos\left(\frac{2\pi\,t}{\tau}\right), & 0 < t \leq \tau/2, \\
\sin\left(\frac{2\pi\,t}{\tau}\right), & \tau/2 < t \leq \tau.
\end{cases}
\]

\noindent where \(\lambda\) is a constant introduced to ensure that \(g_i(t)\) is dimensionless. The coefficients \(a_n\), \(b_n\), \(c_n\), and \(d_n\) are given below.

\noindent For constant driving:

\[
a_0 = c_0 = 1, \quad a_n = c_n = 0, \quad b_n = -d_n = \frac{-1 + (-1)^n}{\pi n}.
\]

\noindent For linear driving:
\[
a_0 = c_0 = \frac{\tau}{4}, \quad 
a_n = -c_n = \frac{((-1)^n - 1)\tau}{2\pi^2 n^2},\]
\[ 
d_n = -\frac{\tau}{2\pi n}, \quad 
b_n = (-1)^n d_n.
\]

\noindent For harmonic driving:
\[
a_1 = d_1 = \frac{1}{2}, \quad 
a_n = d_n = 0 \quad \forall n \neq 1, \quad 
b_1 = c_1 = 0,
\]
\[
b_n = \frac{1 + (-1)^n}{(n^2 - 1)\pi} \cdot n = n \cdot c_n \quad \forall n \neq 1, \quad 
c_0 = -\frac{2}{\pi}.
\]

\noindent By inserting these coefficients into the expressions for \(a_{1vk}\), \(a_{2vk}\), \(b_{1vk}\), and \(b_{2vk}\), the Onsager coefficients can be directly evaluated.

\section{Main expressions for general overdamped Brownian Engines}
\label{appendixover}

Following Refs. \cite{filho1,mamede2022,harunariasy,angel,Noa1}, here we briefly outline the key features of the overdamped regime. Consider a Brownian particle with mass \(m\) coupled to a thermal bath at temperature \(T_i\), whose dynamics are governed by the Langevin equation

\begin{equation}
    \frac{dv_i}{dt} = -\frac{\alpha}{m} v_i + F_i(t) + \xi_i(t),
\end{equation}

\vspace{1mm}

\noindent This equation can be seen as formally equivalent to that of an overdamped harmonic oscillator subject to a force \( f_i(x) = -\kappa x_i / m \). This equivalence becomes evident by replacing \(x\to v\), \( \kappa / \alpha \rightarrow \gamma \) and \( 1 / \alpha \rightarrow \gamma / m \). Thermodynamic quantities such as work, heat, and entropy production are derived analogously to the underdamped case, using the corresponding Fokker-Planck equation. In this framework, the time evolution of the probability distribution \(P_i(v,t)\) follows

\begin{equation}
    \frac{\partial P_i}{\partial t} = -\frac{\partial J_i}{\partial v} - F_i(t) \frac{\partial P_i}{\partial v}
\end{equation}

\noindent where \(J_i\) represents the probability current given by

\begin{equation}
    J_i = -\gamma v P_i - \frac{\gamma\,k_B\,T_i}{m} \frac{\partial P_i}{\partial v},
\end{equation}

\noindent which matches Eq.~(\ref{currentsFK}) in the main text. In the NESS, the mean energy evolves as \(U_i(t) = \frac{1}{2} m \langle v_i^2 \rangle\), and its time derivative can be decomposed into two contributions identical to those presented in the main text. The same applies to the entropy production. Averaging the instantaneous power \(\dot{W}_i(t)\) over a full cycle yields relations similar to Eqs.~(\ref{w11}) and (\ref{power}), whose Onsager coefficients take the forms found in Refs.~\cite{angel,Noa1}:

\begin{equation}
    L_{11} = 1 - \frac{2}{\gamma \tau} \tanh\left(\frac{\gamma \tau}{4}\right),
\end{equation}

\begin{equation}
    L_{12} = \frac{2}{\gamma \tau} \tanh\left(\frac{\gamma \tau}{4}\right),
\end{equation}

\noindent for constant drivings, where \(L_{11} = L_{22}\) and \(L_{12} = L_{21}\), and

\begin{equation}
    L_{11} = \frac{1}{12 \gamma \tau} \left( \gamma^3 \tau^3 - 3(\gamma^2 \tau^2 - 8) \coth\left(\frac{\gamma \tau}{2}\right) - 24 \, \mathrm{csch}\left(\frac{\gamma \tau}{2}\right) \right),
\end{equation}

\begin{equation}
    L_{12} = \frac{(-\gamma \tau + 2 e^{\frac{\gamma \tau}{2}} - 2)\left(e^{\frac{\gamma \tau}{2}}(\gamma \tau - 2) + 2\right)}{2 \gamma \tau (e^{\gamma \tau} - 1)},
\end{equation}

\noindent for linear drivings, where again \(L_{22} = L_{11}\) and \(L_{12} = L_{21}\).

\bibliography{refs}

\end{document}